\documentclass[submission]{eptcs}
\usepackage[T1]{fontenc}
\usepackage{graphicx}
\usepackage{hyperref}
\usepackage{color}
\usepackage{paralist}
\usepackage{xspace}

\newcommand{\FBlt}{\ensuremath{<_{\null_{\mathsf{FB}}}}}
\providecommand{\event}{FROM 2026} 

\newcommand{\intermediate}[1]{\texttt{#1}\xspace}
\newcommand{\Rocq}{Rocq\xspace}
\newcommand{\CompCert}{\textsc{CompCert}\xspace}
\newcommand{\Chamois}{\textsc{Chamois-CompCert}\xspace}
\newcommand{\CompCertELF}{\textsc{CompCert-ELF}\xspace}
\newcommand{\mcpy}{\textrm{Memcpy}\xspace}
\newcommand{\RTL}{\intermediate{RTL}}
\newcommand{\BTL}{\intermediate{BTL}}
\newcommand{\OCaml}{\textsc{OCaml}\xspace}
\newcommand{\GCC}{\textsc{GCC}\xspace}
\newcommand{\Clang}{\textsc{Clang}\xspace}

\begin{document}

\title{Certified Compilation in the TELEPERM XS Nuclear Safety I\&C Platform}
\author{Alexandre Berard
\institute{Framatome SAS Grenoble, 23 Chemin du Vieux Chêne, 38240 Meylan, France}
\email{alexandre.berard@framatome.com}
\institute{Université Grenoble Alpes - CNRS - Grenoble INP - Verimag, France}
\email{alexandre.berard2@univ-grenoble-alpes.fr}
\and
Richard B. Kreckel
\institute{Framatome GmbH, Paul-Gossen-Str. 100, Erlangen, Germany}
\email{richard.kreckel@framatome.com}}

\def\authorrunning{Alexandre Berard and Richard B. Kreckel}
\def\titlerunning{Certified Compilation in the TELEPERM XS Nuclear Safety I\&C Platform}

\maketitle

\begin{abstract}
The large safety instrumentation \& control (I\&C) systems in civil nuclear
power plants (NPPs) are mainly safe-shutdown systems (reactor protection) or
limitation and control systems. Framatome's established TELEPERM XS (TXS Core)
product family is a digital I\&C system platform to cover all these
applications. We illustrate the role of verification in the different stages of
the software production toolchain, focus on the formal compilation process, and
discuss the contribution of the \CompCert certified compiler to the safety case
of the product. Scrutinizing the object code produced by this compiler has
exhibited suboptimal run-time performance in a certain simple but recurring
generated code pattern. We explain how formal methods allow us to address this
issue in the compiler while simultaneously reducing its trusted computing base
(TCB), thereby strengthening the safety case rather than merely preserving it.
\end{abstract}

\section{Introduction}

I\&C systems important to safety in NPPs are subject to strict requirements on
their specification, development process, and verification \& validation
(V\&V). For example, most national safety authorities require the application
of the IEC codes and standards under the umbrella of
IEC~61513~\cite{iec_615132011_nuclear_2011}. Requirements for software-based
systems are specifically elaborated in the
IEC~60880~\cite{iec_608802006_instrumentation_2006}.\footnote{The IEC~61513
  standard is for nuclear installations what IEC~26262 is for automotive
sector. Both nuclear standards IEC~61513 and 60880 are currently being
updated.}

While that régime steers clear of requiring specific programming paradigms, it
does impose certain constraints for the most safety-critical functions like
the absence of recursion or dynamic heap-based memory. This may help explain
why function block diagrams (FBDs) like those proposed
in~\cite{iec_61131-32025_programmable_2025} continue to be the prevailing
programming paradigm for reactor protection, limitation, and control.

\begin{figure}[t]
    \includegraphics[width=\textwidth]{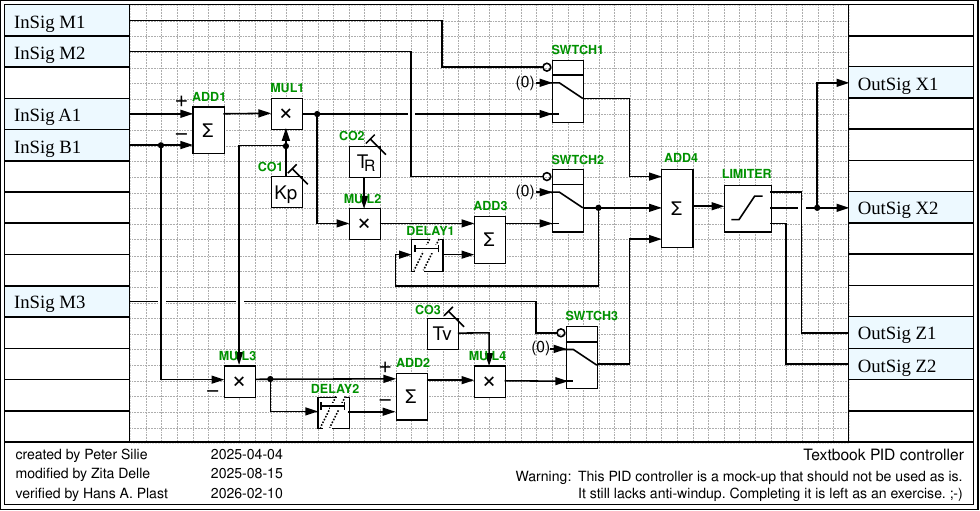}
    \caption{A function block diagram (FBD) implementing a simple standard
      form PID controller. Two analog input signals A1 and B1 and three binary
      active-high `disable' signals M1\dots M3 are interconnected with 17
      function blocks and 24 internal signals to produce the four output
      signals X1, X2, Z1, and Z2.}\label{fig:fbd}
\end{figure}

Systems based on FBDs offer a range of advantages. Their dataflow-based
structure conveniently caters to the process-centric requirements and the
perspective of control engineers. It provides inherent safety from a range of
programming errors. The availability of a pre-developed and fully qualified
function block (FB) library disburdens the control engineers from low-level
programming tasks. Composite FBs (CFBs)~\cite{iec_614992007_function_2007}
suggest themselves as a natural application of the principles of
component-based software development from specification of I\&C functions over
their implementation all the way to V\&V
activities~\cite{zhang_specification_2005}. These advantages are well known and
are described in textbooks, e.g.~\cite{halang_sicherheitsgerichtete_2013}.

The nuclear IEC standards also impose requirements on toolchains for
translating FBDs into code for execution on programmable logic controllers
(PLCs) performing safety functions in NPPs. There is broad
consensus~\cite{v_belgium_licensing_2010} that in addition to comprehensive
validation of the final result the stages of a qualified toolchain for
producing nuclear safety application software must
\begin{inparaenum}[(i)]
\item produce output that is then {\em comprehensively verified} using diverse
  techniques and
\item provide convincing evidence that they {\em preserve the semantics} of the
  input in the produced output.
\end{inparaenum}
In this vein, chapter 14 of IEC~60880 requires mitigation against errors which
could potentially be introduced by translation tools.

This paper is organized as follows: In Section \ref{sec_TXStoolchain} we
introduce the two stages of Framatome's TXS Core
toolchain~\cite{siemens_power_corporation_teleperm_2000,a_graf_appendix_2004,iaea_annex_2018}:
Verified code generation and certified compilation. The second stage is based
on the \CompCert\footnote{\CompCert's official sources:
\url{https://github.com/AbsInt/CompCert}.} verified compiler. We will show why
the first stage naturally produces certain repetitive patterns of code. In
section \ref{sec_Alexandre} we will present work which aims at improving the
executable code compiled from these patterns. This obviously involves the
compiler itself, but also turns out to require a small re-organization of the
code generation.  Section \ref{sec_Related} overviews related work and in
section \ref{sec_Conclusion} we will put this into the perspective of the
general safety case.

\section{The TXS Core Fully Verified Toolchain}\label{sec_TXStoolchain}

At the user level, Framatome's TXS Core toolchain\footnote{To be precise, we
describe the TXS Core 4.5 platform. Earlier versions differ in several
details.} begins with a graphical editor: I\&C engineers drag graphical FBs
from a menu of over 100 different pre-defined and qualified FB types and
interconnect them with signals representing the dataflow between the FBs and
I/O. This results in FBDs or in CFB types which can then be instantiated on
FBDs. I/O comes in several forms: It is either electrical (4\dots20\,mA analog
signals, 24\,V binary signals, etc.) driven by dedicated I/O modules
communicating with the PLC over a backplane bus or it consists of digital
signals transmitted in network messages using a deterministic protocol.
Fig.~\ref{fig:fbd} shows a FBD which could alternatively be the implementation
of a CFB.

\begin{figure}[t]
  \centering \includegraphics[width=\textwidth]{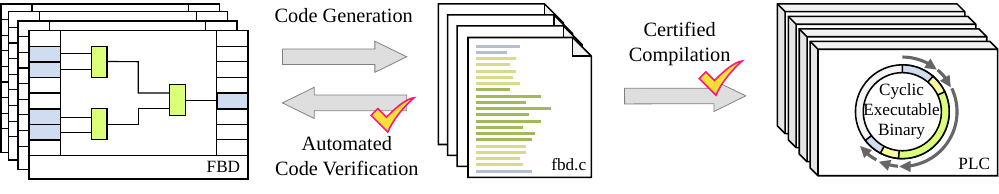}
  \caption{The TXS Core toolchain translates FBDs into certified binaries.}\label{fig:toolchain}
\end{figure}

The subsequent stages (see Fig.~\ref{fig:toolchain}) are fully automated. No
user intervention whatsoever is required (in fact, not even permitted). Below,
we will unfold them.

\subsection{Code Generation}\label{sec_CG}

Using the persisted graphical FBDs as input, a code generator first produces
ISO C code. Largely, it consists of function calls to the standardized,
qualified, and pre-compiled FBs and of glue code representing the dataflow
between the blocks. This ISO C code is often long but always linearly
structured and is itself suitable for a spectrum of analyses.\\

\noindent\textbf{A note on the use of C:} ISO C has proven to be a rewarding
intermediate interface. There is an international
standard~\cite{isoiec_98992024_information_2024}. When restricted to a subset,
the language's ambiguities can be fully avoided. There exists an ample
toolscape for performing static analyses (Astrée, FramaC, etc). Last, but not
least, it is the lingua franca for interfacing with the gamut of commercial
simulator platforms so that the entire I\&C application can be close-looped
with plant models for all purposes from full-scope simulation to operator
training~\cite{richter_verification_2003}.\\

Nevertheless, translation from FBDs to ISO C code requires careful
verification to rule out any error in the code generator.

\subsection{Automated Code Verification}\label{ref_CGverif}

In order to ensure the correctness of the translation from the FBD
specifications to ISO C code, Framatome has developed a fully automated
independent verifier operating on the persisted graphical FBDs and the
generated code as input.  This verifier exploits the fact that an FBD is a
directed acyclic graph of edges connecting FB ports (the netlist).  Every
signal line connecting two FBs is equivalent to the constraint of causality:
The source FB\textsubscript{src} must be invoked before the destination
FB\textsubscript{dst} within each PLC cycle.  In the generated code,
causality is equivalent to the constraint that the source
FB\textsubscript{src} must be called before the destination
FB\textsubscript{dst}.

For example, in Fig.~\ref{fig:fbd}, FB instance {\small\sf ADD1} must be
called before FB instance {\small\sf MUL1}, symbolically {\small\sf ADD1}
\FBlt\ {\small\sf MUL1}. We also find {\small\sf CO1} \FBlt\ {\small\sf MUL1},
{\small\sf MUL1} \FBlt\ {\small\sf SWITCH1}, and by transitivity {\small\sf
ADD1} \FBlt\ {\small\sf SWITCH1}, etc. On the other hand, there is no
causality relation between blocks {\small\sf ADD1} and {\small\sf CO1}.
Obviously, the interconnections on the FBD induce a strict weak ordering
between all FB instances.

A special case is the integration loop between FB instances {\small\sf ADD3},
{\small\sf SWITCH2}, and {\small\sf DELAY1} in Fig.~\ref{fig:fbd}: On first
glance, this creates a strongly connected component which violates the
ordering. Ordering is re-established by breaking the {\small\sf DELAY}-type
block into two parts where the second part is {\em not} causally related to
the first part within the same cycle (but in the next cycle).

This strict weak ordering of FBs suggests an efficient node-by-node and
edge-by-edge method for the translation validation of the generated ISO C code.
A parser reads the generated code and performs the following program:
\begin{itemize}
\item Construct the list of all FB instances, attributed by their types and,
  if applicable, all other artifacts like the FBs' parameters, etc.
  \begin{itemize}
  \item Verify that each element in this list has an exact counterpart in the
    specified FBDs and that no other FB instances than the ones in this list
    are specified.
  \end{itemize}
\item Construct the set of all I/O drivers and network messages, together with
  their counterparts (I/O module, communication partner) and the lists of
  signals they contain.
  \begin{itemize}
  \item Verify that each element of this set has an exact counterpart in the
    specification and no other I/O drivers or network messages are specified.
  \end{itemize}
\item Construct the `netlist' of all interconnections on the FBDs. The
  interconnections are attributed by signal types, possible signal negations,
  their sources and destinations.
  \begin{itemize}
  \item Verify for each signal in this netlist that it is connected with
    source and destination(s) as specified in the specification FBD, that
    their attributes match, and that no other connections exist in the FBD.
  \item For each signal in the netlist, verify that the ordering relation
    \FBlt\ of its source and destination(s) in the generated code complies
    with the one in the FBD.
  \end{itemize}
\end{itemize}
Note the softness of the last item: One can only verify causality for all FB
instances. This is a necessary and sufficient condition for the translation's
semantic preservation. It would not be possible to compare the sequentialized
list of FBs in the generated code with a list of FBs in the specification
because the latter list is not unique. After all, the FBD contains many FBs
which are incomparable w.r.t. \FBlt\ (like {\small\sf ADD1} and {\small\sf CO1}
in Fig.~\ref{fig:fbd}).

Successfully performing the above program proves that the semantics of the FBD
is indeed preserved in the generated application software: the structures are
the same and the generated code respects the FBD's implicit causal relations.

An important overarching non-functional safety requirement on the automatic
verifier is that its implementation must be fully diverse from the code
generator and not share any source code with it. Without this independence, it
would be possible for the two tools to share a hidden fault leading to the
non-detection of an error in the generated code.

Only when the verifier has successfully terminated, the compiler will be
invoked on the generated ISO C code.

\subsection{Certified Compilation}

In a second stage, the generated ISO C code is compiled to ArmV7 object code by
the \CompCert certified compiler~\cite{leroy_formally_2009}.
The object files can then be linked with any of several ELF linkers. \CompCert
performs a series of optimizing transformations based on internal intermediate
representations. These transformations are fully formally proven to preserve
the semantics. This proof is formalized and automatically checked by the \Rocq
Prover (formerly Coq) each time \CompCert is built from sources.

Although formally rigorous, \CompCert's proof of correctness is limited by its
Trusted Computing Base (TCB), which is extensively studied
in~\cite{monniaux_trusted_2022} and mentioned in sections \ref{sec_suboptimal}
and \ref{sec_improvements}. Within TXS Core, the use of \CompCert extensions
which are not proven correct is strictly avoided.

Beyond that, studies subjecting compilers to synthetically generated C code
have independently confirmed that rate of bugs in \CompCert is orders of
magnitude below that of usual compilers~\cite{yang_finding_2011}.

\section{\CompCert Improvements}\label{sec_Alexandre}

The generated ISO C code compiled by \CompCert benefits from its safe proof of
semantic preservation, but it suffers from its lack of some state-of-the-art
code optimizing algorithms that many other compilers have. In fact, the
introduction of such algorithms must be followed with a proof of their semantic
preservation, which require significant work, and may require changing some
aspects of \CompCert's theory models. On top of \CompCert's already existing
algorithms such as dead code elimination and constant propagation,
\Chamois\footnote{\Chamois's official sources:
\url{https://gricad-gitlab.univ-grenoble-alpes.fr/certicompil/Chamois-CompCert}},
a frequently updated fork of \CompCert's official releases, features several
more aggressive code optimizations and uses a framework that facilitates their
writing.

The following sections will show how both \CompCert and \Chamois fail to
optimize TXS Core's generated ISO C code and explains how our recent \Chamois
improvements solve this issue.

\subsection{Suboptimal Patterns Identified in Generated Code}\label{sec_suboptimal}

\begin{figure}[t]
  \includegraphics[width=\textwidth]{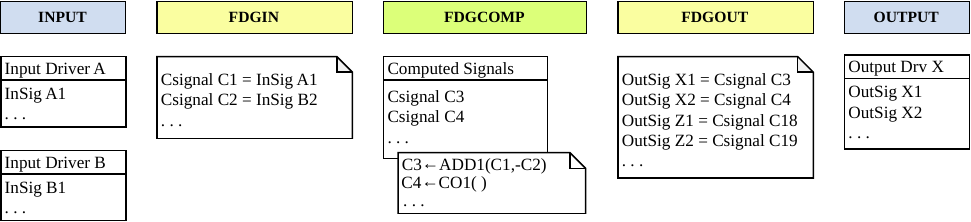}
  \caption{Data moves for computing the FBD from Fig.~\ref{fig:fbd} in the TXS
    Core cycle. The FDGIN (FDGOUT) function routes input (output) signals from
    one structure to another one.} \label{fig:sigassignments}
\end{figure}

The code generated from FBDs contains long repetitive patterns of assignments,
essentially present in places where we route I/O signals towards and away from
compute structures. Those can be seen in Fig.~\ref{fig:sigassignments},
especially in the FDGIN and FDGOUT functions. Also, each signal is translated
into a small C structure, causing the compiler to translate the said
assignments into heavy structure copies.

This is where we observe a weakness: in \CompCert semantics, a structure copy
is translated into a compiler built-in function dedicated to copying data,
which has valid semantics, but needs to be expanded.  And this built-in
function, abbreviated here with \mcpy (not the C libary \texttt{memcpy}
function), is immutably propagated throughout all the intermediate
representations of code, all the way before the back-end part of the compiler
dedicated to emitting assembly code. Such built-in functions are not sensitive
to any code optimization, which explains why these patterns that extensively
use \mcpy are poorly compiled. Moreover, their expansion to assembly code is
directly performed by \OCaml code with no verification; this code belongs to
the TCB of \CompCert~\cite{monniaux_trusted_2022}. Causing any change to this
unproven code is not recommended, especially if we want to apply complex and
unproven optimizations over it. \Chamois is helpless here as well: currently,
none of its optimizations improves the generation of built-in functions.

\subsection{Overview of \Chamois's Formal Verification Framework}

\Chamois possess an Intermediate language Representation (IR), called Block
Transfer Language (\BTL), plugged to \CompCert's Register Transfer Language
(\RTL) IR. \BTL is provided with a symbolic execution engine, here dedicated to
verify middle-end optimizations. This engine contributes to \Chamois exclusive
block verification framework, whose task is to defensively validate code
optimizing oracles~\cite{gourdin_formally_2023}, written in \OCaml. Sometimes,
the symbolic checker needs some information computed by the oracles to be able
to validate these code transformations: this is part of \Chamois Formally
Verified Framework, proven to be a rigorous block verification
mechanism~\cite{boulme_formally_2021}. This allows the rewriting of operations
computed by invariants and memory operations with pointer nonaliasing, allowing
e.g. the simplification of operations across several blocks, or the permutation
of load and store instructions. Similarly, some back-end optimizations are
verified with another symbolic execution engine, but over an abstracted
representation of basic blocks of generic assembly operations, obtained by
proven translation from the configured target architecture. The latter allows
the writing of a postpass instruction scheduling algorithm, where instructions
can also be mutated (\emph{peephole optimization}), which we ported to the
ArmV7 backend. For our case, validating \mcpy transformations required us to
make use of all the cited features.

\subsection{Improvements Made to Compiler Built-in Memory Copies}\label{sec_improvements}

\RTL and \BTL place \mcpy in an algebraic type dedicated to compiler built-in
instructions. \mcpy semantics are simple: it consists in
\begin{inparaenum}[(i)]
  \item loading contiguous bytes from a source memory location, then
  \item storing those bytes to a destination location,
\end{inparaenum}
given the amount of data to copy and memory alignment.

\begin{figure}[t]
  \centering
  \includegraphics[width=0.9\textwidth]{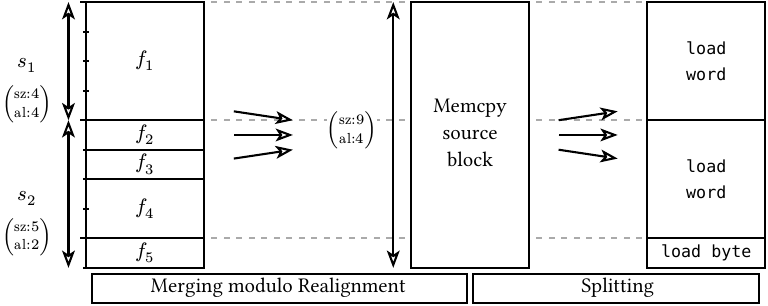}
\caption{Abstracted representation of our \mcpy manipulations, by viewing only
  at source addresses of structures in-memory (layout is assumed to be similar
  for destination addresses). The two structures, $s_1$ and $s_2$, are merged
  and realigned (left part), and so, can be fragmented into a series of two
word-sized and one byte-sized loads (right part). The size (sz) and alignment
(al) is also specified for each structure, as well as their fields (e.g. $f_2$
to $f_5$ for $s_2$) of varying sizes.} \label{fig:memcpy}
\end{figure}

Given the form of suboptimal patterns observed in section \ref{sec_suboptimal},
we came up with ideas that would improve \CompCert's \mcpy. We first found that
the offset of each field of a structure is statically known, as they are placed
in the same memory block. This made us think about a way to \emph{merge} memory
copies where both the source and destination offsets are adjacent. Then,
another idea arose: since the alignment of the global memory block is also
statically known, the internal structural copies could benefit of a
\emph{realignment} analysis, that we use to determine a better width of copy
instruction to use. A picture of these ideas is shown in Fig.~\ref{fig:memcpy}.
The addition and usage of realignment analysis conducted us to change several
aspects of \CompCert's inner memory model~\cite{berard_formally_2026}. The
following paragraphs will briefly explain how we incorporated these new ideas
within \Chamois.

\subsubsection{Merging memory copies with a verified oracle}

This part aims at manipulating several \mcpy instructions together, merging
those which have adjacent copy addresses. We opt to do this part at the \BTL
stage, benefiting from its symbolic execution engine capable of validating
memory transformations obtained from an untrusted oracle. Using this principle,
we can write a merger oracle that can reorder and simplify instructions --
for instance, minimizing the number of virtual registers used in \mcpy
addressing arguments, thereby reducing the number of blocks that may differ
from one to another. The symbolic checker also supports validating the
reordering of memory operations, including \mcpy, through alias
analysis~\cite{berard_formally_2026}, which further increases opportunities for
merging.

The merger oracle consists in greedily finding valid \mcpy candidates for merging.
For this, we reuse the dependency graph from \Chamois instruction scheduling;
we prioritize other instructions and keep the encountered \mcpy instructions
in a separate set. Once no further instructions can be scheduled, we consider
the remaining set of \mcpy instructions and determine whether any of them can
be merged. Two \mcpy can be merged only if their memory blocks are contiguous,
for both destination and source addresses. Also, we consider the case when
merging two copies cause an overlap between the source and the destination
blocks. This case made us change the semantics of memory copies, where a
writing mode is now specified: \emph{forward} or \emph{backward}.\footnote{This
can be seen as analogous to the \texttt{memcpy} and \texttt{memmove} functions
in the C standard library.} Thus, a copy merge is valid when both copies have
the same writing mode, which must be correct with regard to the overlapping
case. Although the theoretical foundations for validating memory copy merges
via symbolic execution have been fully established, the current oracle does not
handle overlapping cases; this is left for future work.

The symbolic checker also needs to be adapted. Originally, it was possible in
\Chamois to replace two single word store instructions with a double word one
during peephole selection~\cite{six_formally_2022}. We extend the same
principle for the validation of merged \mcpy, despite the need for further
information. Through the oracle, we annotate each merged \mcpy with a list of
all the original candidates. This information is used at the rewriting stage in
order to validate merges with regard to the source code, relating for instance
the address ranges and rewriting modes of each copy from the source with the
copy list. The generated annotation must also exhibit the same behavior than
the \mcpy it is attached to. Note that the annotation list does not necessarily
need to preserve the source order, since the symbolic checker can deal with the
reordering of memory operations. For proof simplicity, we still have to provide
the copies in increasing order of address offsets in \emph{forward} copy mode,
and vice versa for the \emph{backward} mode.

\subsubsection{Realignment analysis}

Increasing the width of selected copy instructions used for memory copies, as
shown in Fig.~\ref{fig:memcpy} would obviously decrease the total number of
load and store instructions after expansion. This was done by retaining the
global alignment of the memory block as $\beta$, which could be a power of $2
\in \{1,2,4,8\}$ in bytes. For global variables, it corresponds to the maximum
alignment among all fields in the memory block.\footnote{This is the alignment
of global variables that \CompCert provides to the linker.} The computation
happens during constant propagation where more information about \mcpy
alignment is propagated: formerly only retaining its size and alignment, we now
also registers the couple $(o, \alpha)$, where $\alpha$ is the maximal
alignment that can be set, and $o$ being the necessary offset to reach that
maximal alignment. This pair is computed with respect to $\beta$ and the global
source and destination offsets of a \mcpy address within the block, under
various case-by-case scenarios.

As an example, from Fig.~\ref{fig:memcpy}, realignment is performed on
structure $s_2$; it is not shown on the diagram, but the realignment analysis
propagates $\alpha = 4$ from the previous block ($s_1$) where $al=4$. This step
is necessary for merging blocks $s_1$ and $s_2$ into the bigger block with
$al=4$. Offsets are also necessary, e.g. for a copy of fields from $f_3$ to
$f_5$, assuming a pointer cast to a structure of valid shape, we can still
derivate $\alpha = 2$ given $o = 1$, i.e. we can reach a best alignement of $2$
after an offset of $1$.

This realignment upgrade allows us to fragment \mcpy into bigger parts at
expansion stage. Also, it integrates quite well with the fusion of \mcpy,
giving the opportunity of using wider copy instructions across several
structures whose assignments were merged.

\subsubsection{Formal expansion of memory copies with fragmentation}\label{sec:expansion}

We wrote a fully proven pass in \BTL that does a formal expansion of \mcpy
instructions. Symbolic execution engines are not involved here: we wrote a
general \emph{local} expansion functor that defines an \emph{expanse} function,
verified with a \emph{local} simulation proof. More precisely, this function
replaces each \BTL block with a sub-Control~Flow~Graph (CFG) of \BTL blocks that
locally simulates the source block. And the function was instantiated for the
expansion of \BTL blocks containing \mcpy. The sub-CFG structure allows the
expansion of loops, which is useful for the expansion of large \mcpy. The
writing and the validation of the \emph{expanse} function was simplified by the
use of a \emph{state+error} monad, which propagates two information: the
\emph{fresh} registers and labels introduced by the expansion. The \emph{local}
simulation proof ensures that the \emph{expanse} function returns a valid
transformation, where the state of execution (register set, memory) must be
preserved, excluding the \emph{fresh} values newly introduced within the
expansion monad.

Expansing a \mcpy instruction consists in fragmenting the copy block in
\emph{atomic} parts that can directly be translated to hardware instructions.
By definition, an \emph{atomic} \mcpy has the same size as its memory
alignment. The fragmentation is not so trivial: with the formerly computed
realignment information, we are able to incrementally select wider sizes for
the leading \emph{atomic} block. Thus, the resulting sequence of fragmented
\mcpy can be decomposed in three parts:
\begin{inparaenum}[(i)]
  \item a \emph{prologue}, containing all the necessary \emph{atomic} blocks
    needed prior to reaching the maximum realignment. It has the same size as
    the offset $o$ computed during realignment analysis.
  \item A \emph{copy part}, containing a repetition of atomic blocks having the
    size of the best alignment $\alpha$, and
  \item an \emph{epilogue}, being the remainder of \mcpy block left to be
    fragmented. An example of \emph{epilogue} is visible in
    Fig.~\ref{fig:memcpy}, which corresponds to the last load of a single byte.
\end{inparaenum}
For large copies, the \emph{copy part} may be done with a loop, since the same
copy instruction is repeated. The loop body has been designed to only have 4
assembly instructions performing the copy:
\begin{inparaenum}[(i)]
  \item a post-incremented load,
  \item a post-incremented store,\footnote{Those instructions, which are too
    architecture-specific to live in operations of \RTL, are selected with
  \Chamois peephole algorithm in later passes.}
  \item a pointer comparison upon copy termination, and
  \item the loop's conditional branch.
\end{inparaenum}

\subsubsection{Reduction of \CompCert's Trusted Computing Base}

By normalizing all \mcpy instructions into \emph{atomic} copies, as explained
in Section \ref{sec:expansion}, we eliminate the need for the complex and
unsafe translation of the general-case \mcpy that previously resided in
\CompCert's TCB. This translation is replaced by a significantly simpler one:
following the same principle as assembly printing, it consists of a one-to-one
translation for each case of \emph{atomic} \mcpy instruction, safely derived
from the fully proven expansion described above. Note that we cannot remove
this translation from the TCB entirely -- for instance, by lowering memory
copies to individual loads and stores -- due to a semantic notion attached to
\mcpy pointer fragments that cannot be bound to a valid byte
type~\cite{berard_formally_2026}.

\subsection{Evaluation Within the TXS Core Toolchain}

\begin{figure}[t]
  \includegraphics[width=\textwidth]{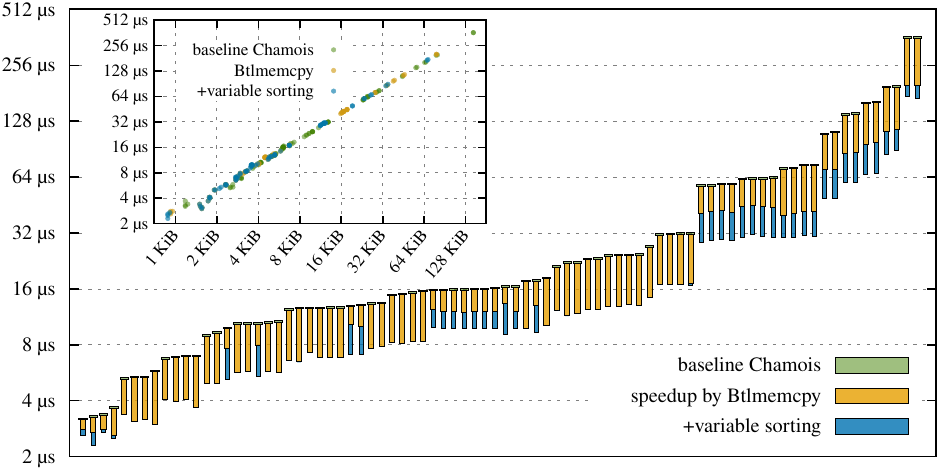}
  \caption{On 82 measured TXS CPUs (arranged horizontally in increasing
    runtime), the \Chamois improvements (Btlmemcpy) speeds up the FDGIN
    functions by a factor of 1.64 on average (beige bars).  Variable sorting
    speeds up by another 1.16 (blue bars). The total average speedup is 1.90.
    The inset compares runtimes with memory footprint of generated and compiled
    code -- it shows that the two are trivially related.}
  \label{fig:bench}
\end{figure}

To assess the effectiveness in a realistic scenario, we set up an experiment
using an actual TXS Core safety I\&C system specification on real TXS Core
hardware.  The results in Fig.~\ref{fig:bench} show that our improvements to
\Chamois significantly speed up the patterns from generated code observed in
section \ref{sec_suboptimal}.

We then went one step further and modified the TXS Core code generator to help
the compiler.  As explained in section \ref{sec_improvements}, we can merge
adjacent memory copies.  However, that highly depends on the layout given in
generated structures.  Especially, the TXS Core code generator was not
initially designed to sort the computed signals such that their (arbitrary)
order matches the (fixed) order in the input structures.  Just sorting these
variables produces more assignments where both the source and the destination
are consecutive.  This additional improvement of 16\% on average is only
effective together with the improvement of the compiler.

\section{Related Work}\label{sec_Related}

A verification method called `diverse back translation' was proposed for
safety-critical code in~\cite{krebs_verfahren_1984}. The authors manually
reconstructed the specification from the implementation and compared it with
the original specification.  They clearly recognized the tedium involved in the
task, even for tiny programs. Much later, it was pointed out
in~\cite{halang_exploiting_1995} that the process could be greatly accelerated
while also obsoleting the need for domain experts performing the task if the
source is a graphical FBD specification.  The \textsc{Retrans}
tool~\cite{miedl_retrans-tool_1998} semi-automates diverse back translation of
generated TXS Core C code.  It includes in its analysis the comparison of
redundant structures -- this goes beyond our verifier. The GET-R1 engineering
tool for the Russian TPTS-SB nuclear safety I\&C platform is similar, but in
the absence of a verified compiler it tackles back translation from bytecode
back to the FBD specification~\cite{belonosov_verification_2018}. Both
\textsc{Retrans} and GET-R1 rely on sequentialization hints from the persisted
specification. This is incomplete because sequentialization is not a priori
part of the specification -- the code generation tool establishes it and often
writes it into the specification as annotations.  As explained in section
\ref{ref_CGverif}, checking the \FBlt\ relation for all edges in the netlist on
the generated code fills this gap.  D.-A. Lee et al. have applied translation
from FBDs {\em including} the FBs to Verilog which then made it possible to
check the equivalence of this Verilog program with the target C program with
HW-CBMC, before the C program was compiled for the
PLC~\cite{lee_systematic_2013}.  The verifier integrated in the TXS Core
toolchain may be the first fully automated, complete, and direct diverse back
translation tool used in the nuclear industry.
A promising approach to formal compilation of synchronous dataflow languages is
the Vélus project~\cite{EMSOFT23}, a \Rocq-written compiler that
translates Lustre programs directly to \CompCert's \emph{Clight}, bypassing C
generation and subsequent code verification. However, it does not yet appear
integrated into a proper tooling environment suitable for compiling graphical
FBDs, as required by nuclear industry standards.

The \CompCert certified compiler has been used before for safety functions in
the nuclear industry~\cite{kastner_compcert_2018}.  There, additional
qualification steps are performed.  Notably, the fully linked executable is
checked using the Valex tool against a serialization of \CompCert's internal
assembly representation to verify that the assembler and linker did not
introduce an error. In the same vein, \CompCertELF is an extension that embeds
this verification at the core of the verified compilation chain, extending the
verification from the assembly file up to the ELF object
file~\cite{wang_compcertelf_2020}.

Implementers have long been optimizing library functions for copying memory
blocks by taking advantage of wide registers and handling alignment. Unverified
compilers like \GCC and \Clang are also able to optimize two
adjacent 4-byte copies by one 8-byte copy. But performing memory copy merging
as discussed in section \ref{sec_Alexandre} seems to be
novel~\cite{berard_formally_2026}. The structures and assignments generated so
naturally for embedded PLC software seem to be have escaped popular benchmarks.

\section{Conclusion and Outlook}\label{sec_Conclusion}

In the nuclear industry, formally verifying that the translation from FBDs to
the executable code preserves the semantics does not and will not obsolete
comprehensive I\&C system validation.  Still, a certified toolchain is crucial
since it justifies the trust placed in model
checking~\cite{lahtinen_model_2016} and simulations. After all, these
disciplines are usually exercised directly on the FBD specifications.

In this work we were able to address a performance weakness of the \CompCert
compiler leveraged by the TXS Core toolchain for translating generated C code.
This led to code which is not only much smaller and faster, but also reduces
the compiler's TCB and this directly improves the safety case. This is very
important: While increasing the trust in a nuclear safety platform's toolchain
is unlikely to obsolete any V\&V activity in test bay, it can significantly
support the introduction of local functional changes to FBDs at a later stage.

Some practical and administrative activities are still needed to introduce our
improvements of \CompCert into the TXS Core toolchain.

\noindent\textbf{Acknowledgements.} The authors would like to thank the
directors of Framatome's I\&C business unit for permission to publish this
paper. R. Kreckel also thanks Prof. W.~Halang for insightful discussions about
diverse back translation and proofs of correctness in general.

\end{document}